\documentclass[preprint,showpacs,preprintnumbers,amsmath,amssymb]{revtex4}

\usepackage{graphicx}
\usepackage{dcolumn}
\usepackage{bm}
\usepackage[usenames,dvipsnames]{color}

\begin{document}

\title{Coherent Control of Tunneling in Double-Well Potentials}

\author{Dae-Yup Song}
\email{dsong@sunchon.ac.kr}
\affiliation{Department of Physics Education,
Sunchon National University, Jeonnam 540-742, Korea}

\date{\today}

\begin{abstract}
For an asymmetric double-well potential system, it is shown that, if
the potential is quadratic until it reaches several times of the
zero-point energies from the bottoms in each well, the energy
eigenvalues of the low lying excited states of the double-well
system must be close to the eigenvalues of the quadratic potentials.
These eigenvalue structures suggest a method for the coherent
control of the tunneling as well as realizing almost complete
localization of the wave packet in one of the wells, by handling the
double-well asymmetry. Numerical examples are included to indicate
that the method could be useful also in a more general potential,
and to propose experimental confirmations.
\end{abstract}

 \maketitle

\section{Introduction}

Quantum tunneling through a barrier is a fundamental physical effect
\cite{AG,Lee}, and the neutral atoms trapped in optical lattices
have given us an opportunity to study the aspects of  tunneling,
including coherent dynamics on macroscopic scale. For the symmetric
double-well system which has long served as a paradigm of quantum
physics, the eigenvalue structures are well-known for the  low lying
excited states \cite{WKB1,WKB2,WKB3,WKB4,GDJH}, and it has been pointed out that, by
adding a specific driving force, the tunneling dynamics can be
brought to a complete standstill known as coherent destruction of
tunneling (CDT) \cite{GDJH}. Recently this coherent destruction has
been visualized in single particle tunneling \cite{KSSTO} as well as
in the tunneling of Bose-Einstein condensates (BECs) \cite{Valle}.
On the other hand, in experiments, localized wave packets have been
prepared in one well of an asymmetric double-well potential, and the
tunneling dynamics has been observed by turning off the asymmetric
part of the potential \cite{HADGJ,DJ,KSSTO}. It is also known that
the density distributions of the BECs of interacting particles are
asymmetric in the asymmetric double-well potentials
\cite{AG,HWAHS,SH,DHC}, to result in the non-vanishing relative
phase evolution rate \cite{JS1,JS2}.

In this article,
we will find the energy eigenvalues of low lying excited states of
an asymmetric double-well potential $V(x)$ which has one local maximum
at $x=x_c$ between the wells, by constructing WKB wave functions with
the quadratic connection formula.
To quantify the degree of tunneling,  we define the  {\em tunneling
visibility} for a wave function $\psi(x,t)$, as ${\cal V}= (P_{max}
-P_{min})/(P_{max} +P_{min}),$ where $P_{max}$ $(P_{min})$ denotes
the maximum (minimum) value of $P_r(t)=\int_{x_c}^\infty \psi^*(x,t)
\psi(x,t) dx$ during the time evolution. We find that the wave
functions of (almost) arbitrary  ${\cal V}$ ranging from 0 to 1 can
be realized from a Gaussian wave packet, by controlling the
potential energy difference between the bottoms of the double well.
The case of ${\cal V}\approx 0$ with $P_{max}\approx 1$ amounts to
the CDT, and this case can be realized when the two wells can be
considered to be separate.
A Gaussian wave packet at a stand still  can be realized also in the
well of higher bottom.

While the results for a single particle can be applied only for
the systems of noninteracting particles, we note that interacting systems
have been extensively studied \cite{MCWW,NK,JI,JMP}. Particularly,
for the systems of a few bosons, highly delayed pair tunneling
analogous to nonlinear self-trapping  has
been found for the medium range of the interaction strength \cite{JMS}.
Though we only consider the double-well potentials bounded
from below, the system of a particle in a periodic potential
with an additional constant force has been of great interest \cite{Sias,BWK,Holthaus},
and we note that this system has been analyzed through the instanton method
\cite{LFZLC}
which is intimately related with the WKB analysis \cite{WKB1}.

In the next section, before presenting the main {\em analytic} results,
two asymmetric systems will be numerically studied.
In Sec.~III, we will construct the WKB wave functions for a general potential
$V(x)$.  It will be shown that the asymmetric systems
can be classified into two different regimes: In the one regime, an eigenfunction
of low lying excited states has
significant amplitude in both wells as in the symmetric systems,
while, in the other, the eigenfunction describes the
particle mostly localized in just one of the wells.
In Sec.~IV, we will develop formulas for the estimation of the energy eigenvalues
in the regime of the localized eigenfunctions, and for the estimation of
the tunneling visibility in the other.
In Sec.~V, the asymmetric double oscillator model
will be numerically solved to give an implication on the eigenvalue structure
of a general double-well system, and to show that {\em WKB description} could be
{\em remarkably accurate}.
The last section will be devoted to a summary and discussions.

\section{Coherent control of Tunneling: Numerical examples}
In this section, we will study two systems numerically to indicate that
the coherent control method could  also be useful in a general potential
whose wells are {\em not} exactly quadratic,
and to expose that gravity may be used to control the tunneling dynamics in
the settings of the recent experiments \cite{KSSTO,HADGJ}.

First,   we
consider the system of a particle of mass $m$ in the quartic
double-well potential
\begin{equation}
V_Q(\alpha;x)=\hbar\omega\left[
-\frac{x^2}{4l_{ho}^2}+\frac{x^4}{96l_{ho}^4}+\frac{\alpha
x}{8\sqrt{3}l_{ho}}+C(\alpha)\right],
\end{equation}
with $l_{ho}=\sqrt{\frac{\hbar}{m\omega}}$, where $C(\alpha)$ is
introduced  to ensure that the minimum of the potential is 0. While
the term proportional to $\alpha$ is added to give the asymmetry,
$V_Q(0;x)$ is a special case of the well-known potentials
\cite{GDJH,WKB1,WKB2,WKB3,WKB4}.
For $V_Q(0;x)$, the angular frequency for small oscillations
at the bottoms of each well is $\omega$, and the barrier height is
$3\hbar\omega/2$.

\begin{figure}
\includegraphics[width=3.3in]{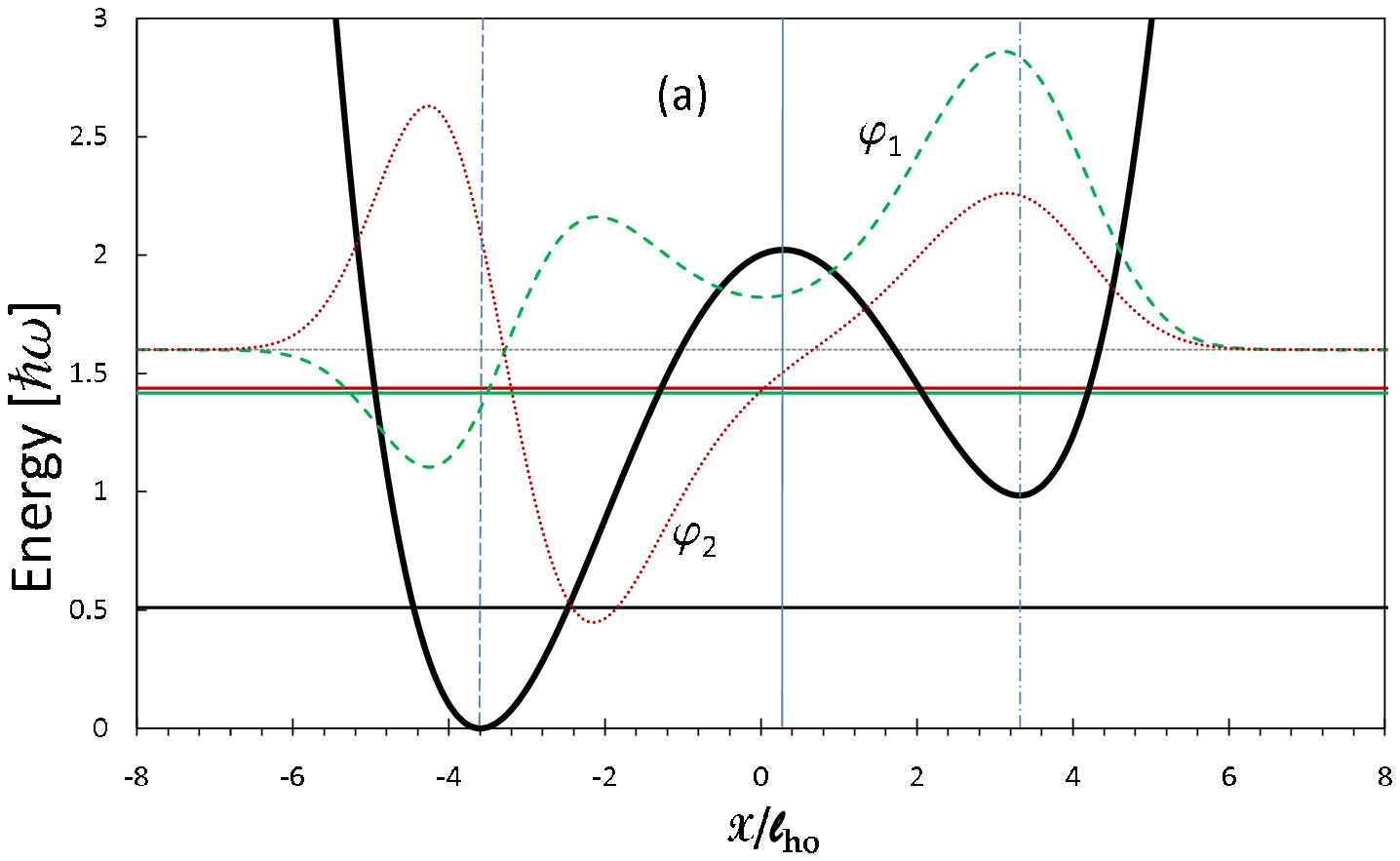}
\includegraphics[width=3.3in]{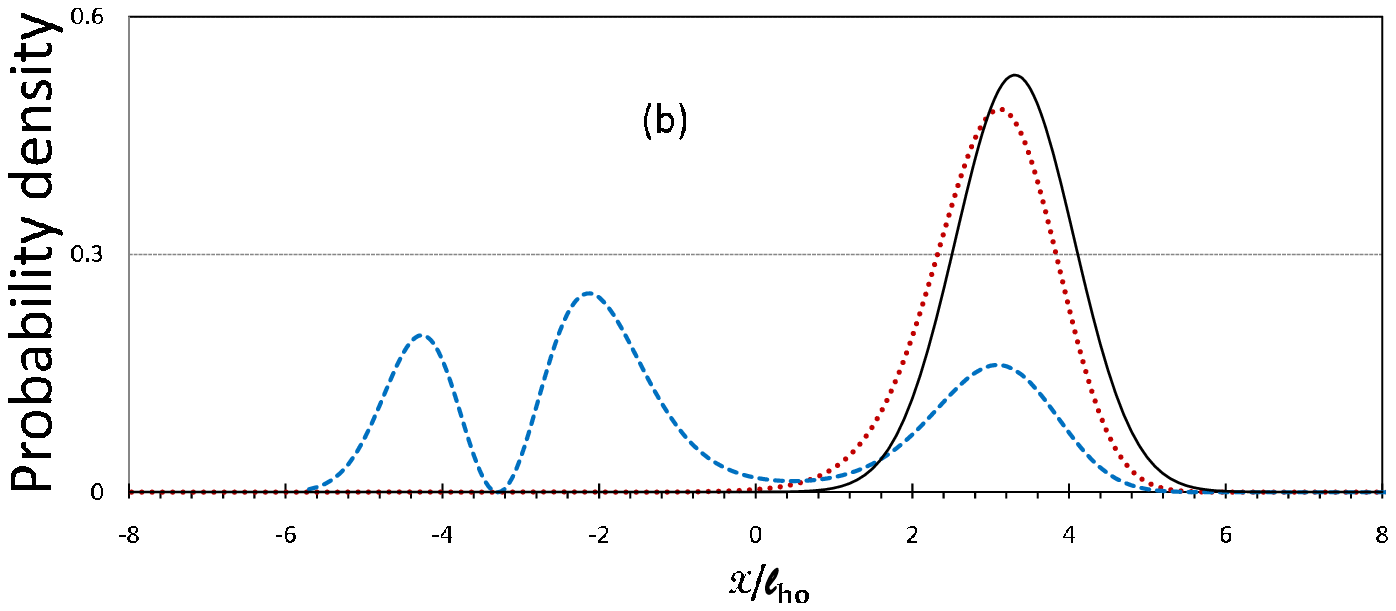}
\includegraphics[width=3.3in]{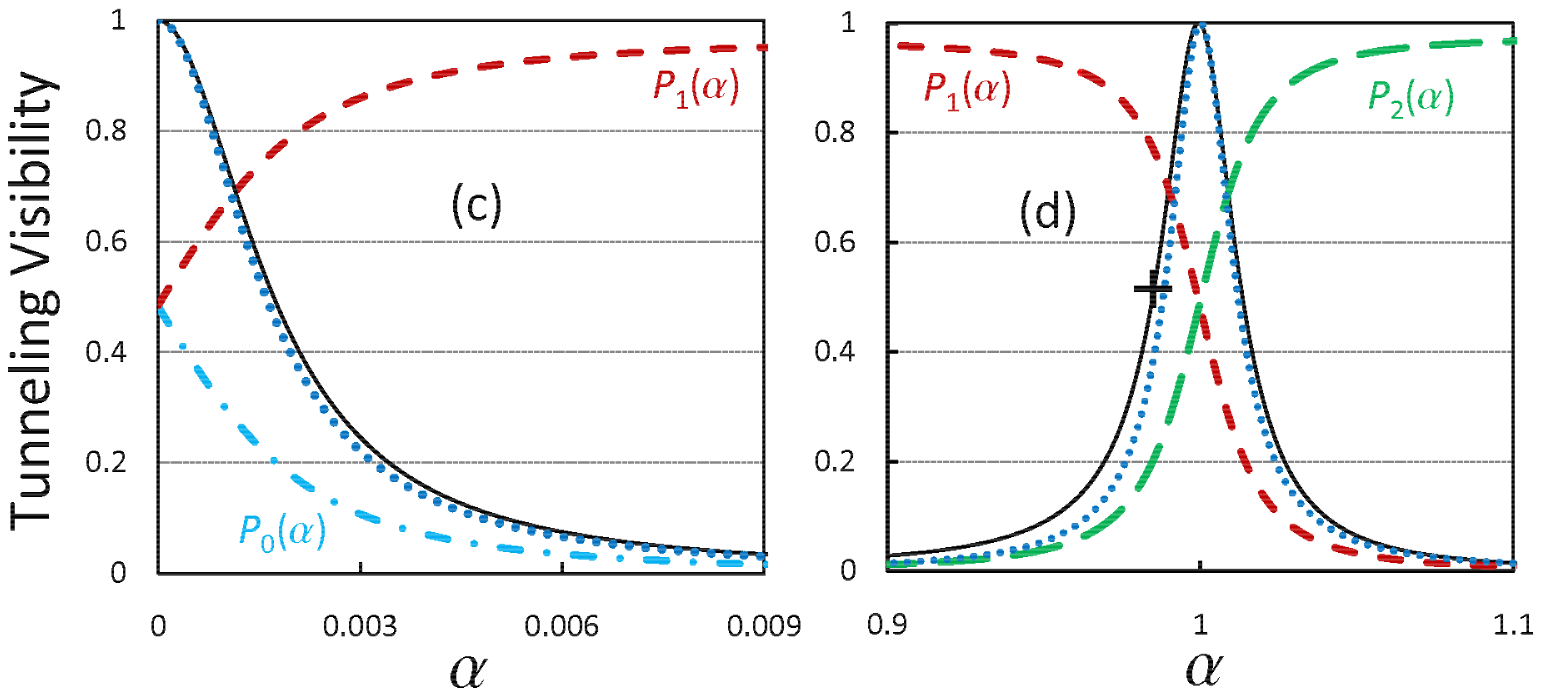}
\includegraphics[width=3.3in]{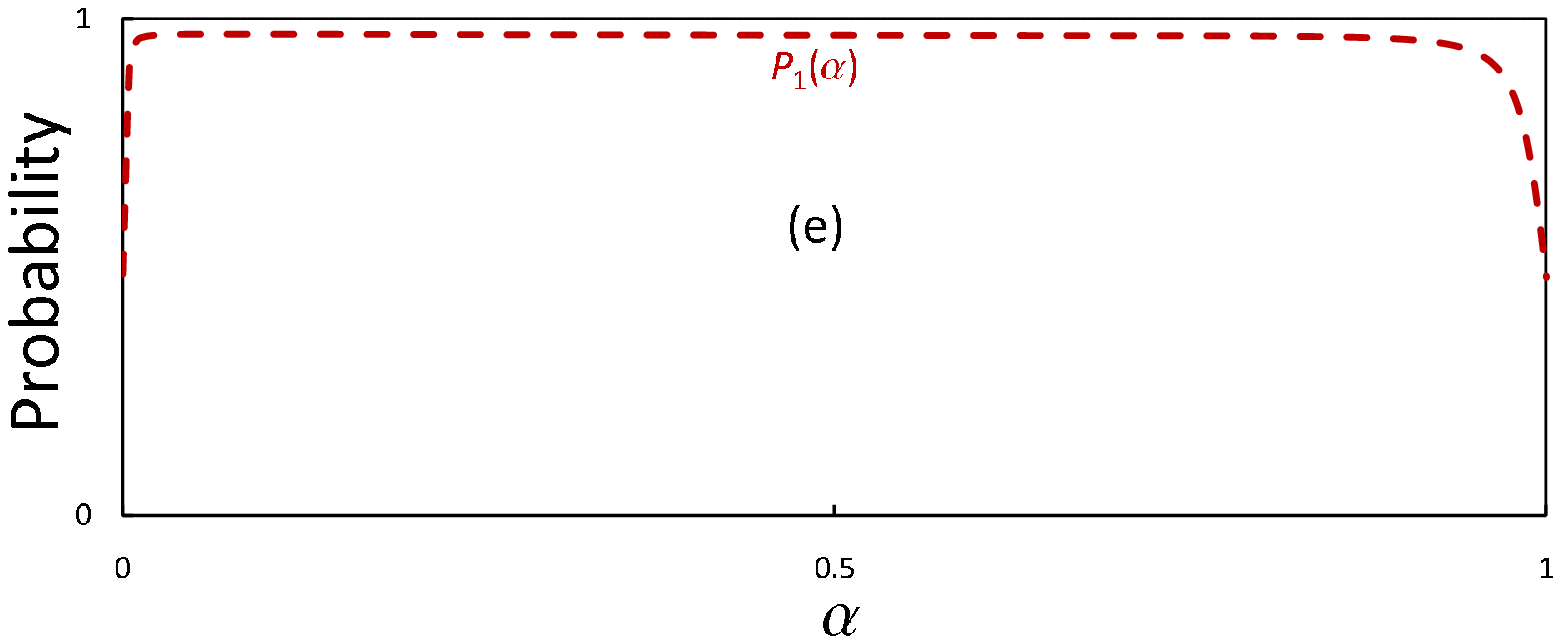}
\caption{(Color online) Tunneling dynamics in $V_Q(\alpha;x)$.  (a)
A potential (thick solid line) and three lowest eigenvalues (thin
black, green, red solid lines), for $\alpha=\alpha_0=0.985$. The
dotted and dashed, dashed, solid vertical lines indicate the values
of  $a(\alpha_0)$,  $-b(\alpha_ 0)$, $x_c(\alpha_0)$, respectively,
where $V_Q(\alpha_0;-b(\alpha_0))=0$. (b) $|\phi_G(\alpha_0;x)|^2$
(solid line), and $|\psi_{\alpha_0}(x,t)|^2$ at $t=0$ (dotted line)
and at $t=\hbar \pi/(E_2(\alpha_0 )-E_1(\alpha_0 ))$ (dashed line).
 (c) and (d)
The calculated tunneling visibility for $\psi_\alpha(x,t)$ (solid
lines) and an estimation (dotted lines). The estimation is made from
Eq.~(\ref{visibility}). For (c),
$\delta_\epsilon/\delta_a=2V_\alpha(a(\alpha))/(E_1(0)-E_0(0)).$ For
(d), since $E_2(\alpha)-E_1(\alpha)$ has the minimum at
$\alpha=\alpha_1=1.00$,
$2[V_\alpha(a(\alpha))-V_{\alpha_1}(a(\alpha_1))]/(E_2(\alpha_1)-E_1(\alpha_1))$
is used as $\delta_\epsilon/\delta_a$.   The "{\bf +}" mark  denotes
the value of $\alpha_0$.  } \label{quartic}
\end{figure}

For the Gaussian wave packet
$\phi_G(\alpha;x)=<x|\phi_G(\alpha)>=\exp[-(x-a(\alpha))^2/(2l_{r}^2)]/(l_r^2\pi)^{1/4}$
centered at the bottom of the right well, $a(\alpha)$, with
$l_r^{-4}=\frac{m^2\omega_r^2(\alpha)}{\hbar^2}=\frac{m}{\hbar^2}\frac{d^2
V_\alpha(x)}{dx^2}|_{x=a(\alpha)}$,  in Fig.~\ref{quartic}, we
evaluate the probabilities ${
P}_i(\alpha)=|<\phi_G(\alpha)|\varphi_i(\alpha)>|^2$, where
$\varphi_i(\alpha;x)=<x|\varphi_i(\alpha)>$ are the eigenfunctions
arranged in the order of ascending energy eigenvalue $E_i(\alpha)$
$(i=0,1,2,\cdots)$. When we define the (unnormalized) wave function $\psi_\alpha (x,t)$ as
\begin{eqnarray}
&\psi_\alpha(x,t)&\cr
&=&\sqrt{P_1(\alpha)}\varphi_1(\alpha;x)\exp\left[\frac{-iE_1(\alpha)t}{\hbar}\right]\cr
&&+\sqrt{P_2(\alpha)}\varphi_2(\alpha;x)\exp\left[\frac{-iE_2(\alpha)t}{\hbar}\right],
\end{eqnarray}
Fig.~\ref{quartic} indicates that, in this system of deep quantum regime,
$\psi_\alpha(x,0)$ closely describes the Gaussian wave function $\phi_G(\alpha;x)$
for $0.01\le \alpha $ in the given range of $\alpha$.
Further, plot (e) shows that, if $ 0.01<\alpha<0.9$,
the Gaussian wave
packet is closely described by an eigenfunction,
$\varphi_1(\alpha;x),$ which is in turn mostly localized in the
right well of the higher bottom (We note that localized eigenfunctions have also
been known in the buried double-well systems \cite{WM}).

As is well-known in the systems of symmetric potentials \cite{WKB1,WKB3},
the wave function $\frac{1}{\sqrt{2}}\left(\varphi_0(0;x)e^{\frac{-iE_0(0)t}{\hbar}}
+\varphi_1(0;x)e^{\frac{-iE_1(0)t}{\hbar}}\right)$ describes the system
which tunnels back and forth between an almost Gaussian state localized
in the left well and the state localized in the right well,
with ${\cal V} \approx 1$.
As, for $0.01\le\alpha\le 0.9$, the ground state wave function $\varphi_0(\alpha;x)$
and the first excited state wave function $\varphi_1(\alpha;x)$
are mostly localized in the left and right wells with the shapes close to a Gaussian, respectively,
if we could change $\alpha$ from 0 to a value
which is larger than 0.01 but smaller than 0.9, without distorting the wave function
by the change,
then we will have a system where the probability density is almost stationary.
For this stationary system, the probability of finding the particle in the left
(or right) well depends on the details of changing $\alpha$.
If the change could be made in a much shorter period of time compared with
the period of tunneling $2\hbar \pi/(E_1(0 )-E_0(0 ))$, the probability of finding the
particle in one of the wells
crucially depends on the timing of the change.
If we tune $\alpha$ back to 0 or to 1, tunneling dynamics appears again,
but in this time the tunneling visibility ${\cal V}$ could be less than 1
depending on the  process.

Second, for an atomic spinor trapped in a double-well of the potential
\begin{eqnarray}
V_L(\beta;x)&=&C E_R[\cos^2 kx +\xi\cos^2
(\frac{kx}{2})-{\xi}/{2}+{\xi^2}/{16}]\cr
&&+m\beta gx,
\end{eqnarray}
with $E_R=(\hbar
k)^2/2m$  \cite{KSSTO,HADGJ}, we explicitly consider the case of
$C=10$ and $\xi=\frac{1}{2}$ which is similar to an experimental
situation \cite{KSSTO}. For $\beta=0$, with the vanishing boundary condition
at $x=\frac{2j\pi}{k}$ ($j:$integer), there are three energy
eigenvalues $E_i^L~(i=0,1,2)$ under the barrier height, with
$E_1^L-E_0^L=0.122E_R$, $E_1^L+E_0^L=5.70E_R$, and $E_2^L=7.27E_R$.

If $\lambda=2\pi/k=811$ nm, $g=980$ ${\rm cm/s^2}$ and
cesium atoms are in $V_L(1;x)$, the fact
$mg\lambda/2=0.580E_R=4.75(E_1^L-E_0^L)$ then suggests that the
tunneling visibility is very low for a cesium atom trapped in the
vertical optical lattice aligned along the Earth's gravity. If the
right minimum of a double-well is located at $x_r$, we construct a
Gaussian wave packet
$\phi(f;x)=<x|\phi(f)>=\exp[-(x-x_r)^2/(2f^2l_{r}^2)]/(f^2l_r^2\pi)^{1/4}$,
here, with  a fitting factor $f$. Indeed, for the first excited state
$|\varphi_1^L>$ of $\beta=1$, we find that
$|<\phi(0.759)|\varphi_1^L>|^2=0.980$ and
$|<\phi(1.00)|\varphi_1^L>|^2=0.945$,
which shows that the eigenfunction
$\varphi_1^L(x)$ is closely described by a Gaussian wave
packet, to prove the low visibility of the wave packet.
As the visibility is high for the symmetric horizonal lattice, this shows that
gravity may be used to control the tunneling dynamics in the optical
lattices.

\section{WKB wave functions}

In this section, we will construct WKB wave functions for a general potential
$V(x)$, assuming that  $V(x)$ is written as $\frac{m\omega_l^2}{2}(x+b)^2$ and as
$\frac{m\omega_r^2}{2}(x-a)^2+\epsilon\hbar\omega_r$ around the
bottoms of the left and right wells, respectively. The eigenvalue structure
will then be found by requiring the WKB wave functions to be asymptotically matched,
in the overlapping regions, onto the exact solutions of the quadratic wells.

In the quadratic
regions of $V(x)$, the eigenfunction is described by the parabolic
cylinder function $D_\eta(z)$, and the eigenfunction of an energy
eigenvalue $\hbar\omega_r (\nu+
\epsilon+\frac{1}{2})[\equiv\hbar\omega_l (\mu +\frac{1}{2})]$ is
written as
\begin{equation}
C_L D_{\mu}\left(- \frac{\sqrt{2}(x+b)}{l_{l}} \right) ~{\rm and}~
C_R D_{\nu}\left( \frac{\sqrt{2}(x-a)}{l_{r}} \right),
\end{equation}
near the bottoms of the left and right wells, respectively,  with
$l_i=\sqrt{\frac{\hbar}{m\omega_i}}$ $(i=l,r)$. On the other hand,
by taking $x_c=0$,
in the region of the barrier we have an approximate solution for the
eigenfunction through the WKB method \cite{WKB1,WKB2,WKB3}, as
\begin{eqnarray}
\psi_{WKB}(x)&=&  \frac{N_R\sqrt{\hbar}}{\sqrt{l_{ho}
p(x)}}\exp\left[\int_0^x\frac{p(y)}{\hbar} dy\right]\ \cr
&&+\frac{N_L\sqrt{\hbar}}{\sqrt{l_{ho}
p(x)}}\exp\left[-\int_0^x\frac{p(y)}{\hbar} dy\right], \label{WKB}
\end{eqnarray}
where
$p(y)=\sqrt{2m[V(y)-(\nu+\epsilon+\frac{1}{2})\hbar\omega_r]}$.  The
relations between the real coefficients $C_L,~C_R,~N_L,~N_R$ may be
given by comparing the eigenfunctions in the regions where the
descriptions by the parabolic cylinder function and by the WKB
function are both valid. On the negative real axis, the asymptotic
expansion of the parabolic cylinder function is \cite{AS}
\begin{eqnarray}
D_\eta(z) &\sim
&\frac{\sqrt{2\pi}}{\Gamma(-\eta)}\exp\left[\frac{z^2}{4}\right]\frac{1}{|z|^{\eta+1}}
\left[1 +O\left(\frac{\eta^2}{z^2}\right)  \right]\cr &&+\cos(\eta
\pi)
\exp[-\frac{z^2}{4}]|z|^\eta\left[1+O\left(\frac{\eta^2}{z^2}\right)\right].
~~~~~ \label{pcf asymp}
\end{eqnarray}

\begin{figure}
\includegraphics[width=3.3in]{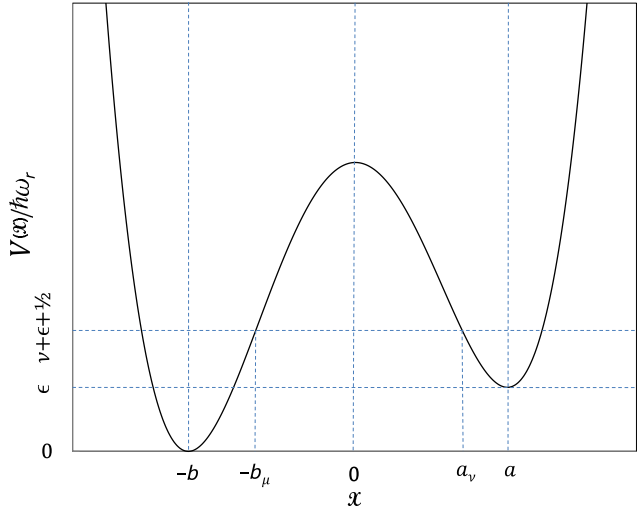}
\caption{(Color online) An illustration of turning points.} \label{vxf}
\end{figure}

The WKB wave function can also be expanded, for instance, as
\begin{eqnarray}
&&\psi_{WKB}(x) \cr && \sim2^\frac{1}{4}N_R
\sqrt{\frac{l_r}{l_{ho}}}\left(\frac{e}{\nu+\frac{1}{2}}\right)
                   ^{\frac{\nu}{2}+\frac{1}{4}}\left(\frac{\sqrt{2}(a-x)}{l_{r}}\right)^\nu
                   \cr
&&~~~~~~~~~~~~~\times
\exp\left({-\frac{(a-x)^2}{2l_{r}^2}+\int_0^{a_\nu}
\frac{p(y)}{\hbar} dy}\right)\cr &&~~~ +2^\frac{1}{4}N_L
\sqrt{\frac{l_r}{l_{ho}}} \left(\frac{\nu+\frac{1}{2}}{e}\right)
                   ^{\frac{\nu}{2}+\frac{1}{4}}\left(\frac{l_{r}}{\sqrt{2}(a-x)}\right)^{\nu+1}
                   \cr
&&~~~~~~~~~~~~~\times \exp\left(\frac{(a-x)^2}{2l_{r}^2}-\int_0^{a_\nu} \frac{p(y)}{\hbar} dy\right),
\label{WKB asymptotic}
\end{eqnarray}
when $\frac{a-x}{a-a_\nu} \gg 1$ and $x$ is in the region of
quadratic potential of the right well,  where $a_\nu, -b_\mu$ are
turning points satisfying
$V(a_\nu)=V(-b_\mu)=(\nu+\epsilon+\frac{1}{2})\hbar\omega_r$ ~~$(a>
a_\nu>0,~b>b_\mu >0)$ (see Fig.~\ref{vxf}). By comparing the leading terms in the
parabolic cylinder function description which originates from the
first term in the right hand side (r.h.s.) of Eq.~(\ref{pcf asymp})
with the relevant terms in the WKB description, we have
\begin{eqnarray}
N_R&=&\frac{\sqrt{\sqrt{2}\pi}}{\Gamma(-\mu)}\sqrt{\frac{\l_{ho}}{l_l}}
             \left(\frac{e}{\mu+\frac{1}{2}}\right)^{\frac{\mu}{2}+\frac{1}{4}}
             e^{\int_{-b_\mu}^0 \frac{p(y)}{\hbar} dy}C_L,\cr
N_L&=&\frac{\sqrt{\sqrt{2}\pi}}{\Gamma(-\nu)}\sqrt{\frac{\l_{ho}}{l_r}}
             \left(\frac{e}{\nu+\frac{1}{2}}\right)^{\frac{\nu}{2}+\frac{1}{4}}
             e^{\int_0^{a_\nu} \frac{p(y)}{\hbar} dy}C_R.
\label{necessary condition}
\end{eqnarray}
If any of $\mu$ or $\nu$ is not close to a non-negative integer,  in
the large separation limit of $e^{\int_{-b_\mu}^0 \frac{p(y)}{\hbar}
dy}\gg 1$ and $e^{\int_0^{a_\nu} \frac{p(y)}{\hbar} dy}\gg 1$,
Eq.~(\ref{necessary condition}) yields
\begin{equation}
|N_R| \gg |C_L| ~~~{\rm and}~~~ |N_L|\gg|C_R|.
\label{noninteger}
\end{equation}
If the WKB condition $p^2(x) \gg \hbar|\frac{dp}{dx}|$ is satisfied,
in the region of the barrier, the first term  in r.h.s.~of
Eq.~(\ref{WKB}) with positive (negative) $N_R$ is a monotonically
increasing (decreasing) function and the second term with positive
(negative)$N_L$ is a monotonically decreasing (increasing) function.
Eq.~(\ref{noninteger}) then implies that if an eigenfunction exist
for such $\nu$, it gives the probability distribution in which the
probability of finding the particle in the barrier region is
considerable. Since $E_\varphi >\int \varphi^*(x) V(x) \varphi(x)$,
if an eigenfunction gives a considerable probability in the barrier
region, the eigenvalue can not be much smaller than $V(0)$.

On the other hand, if any of $\nu$ or $\mu$ is close to a
non-negative integer, the relations in Eq.~(\ref{noninteger}) are
not valid. If $\nu$ is close to a non-negative integer, due to the
singularity in the gamma function, the second term of the r.h.s.~of
Eq.~(\ref{pcf asymp}) can also be a leading term. By comparing this
type of leading term in the parabolic cylinder function description
with the relevant term in the WKB wave function, we have a relation
between $N_R$ and $C_R$. By combining this relation with the first
one in Eq.~(\ref{necessary condition}), we have
\begin{eqnarray}
\frac{C_L}{C_R}&=&\sqrt{\frac{l_l}{l_r}}\frac{\cos(\nu\pi)\Gamma(-\mu)}{\sqrt{2\pi}}
     \left( \frac{\nu+\frac{1}{2}}{e}\right)^{\frac{\nu}{2}+\frac{1}{4}}
     \cr
&&\times   \left(
\frac{\mu+\frac{1}{2}}{e}\right)^{\frac{\mu}{2}+\frac{1}{4}}
\exp\left[- \int_{-b_\mu}^{a_\nu} \frac{p(y)}{\hbar}dy\right].
\label{integer nu}
\end{eqnarray}
If $\mu$ is not close to an integer, Eqs.~(\ref{necessary
condition}) and  (\ref{integer nu}) imply that the eigenfunction
gives the probability distribution in which the particle is mostly
found in the right well.

If $\mu$ is close to a non-negative integer, we have the relation
\begin{eqnarray}
\frac{C_L}{C_R}&=&\sqrt{\frac{l_l}{l_r}}\frac{\sqrt{2\pi}}{\cos(\mu\pi)\Gamma(-\nu)}
     \left( \frac{e}{\nu+\frac{1}{2}}\right)^{\frac{\nu}{2}+\frac{1}{4}}\cr
&&\times \left( \frac{e}{\mu+\frac{1}{2}}\right)^{\frac{\mu}{2}+\frac{1}{4}}
     \exp\left[ \int_{-b_\mu}^{a_\nu} \frac{p(y)}{\hbar}dy\right].
\label{integer mu}
\end{eqnarray}
In this case, if $\nu$ is not close to an integer,
Eqs.~(\ref{necessary condition})  and (\ref{integer mu})  imply that
the eigenfunction gives the probability distribution of the particle
mostly localized in the left well.

For an eigenstate whose eigenvalue is much lower than the barrier
height,  the eigenvalue thus must be close to
$\hbar\omega_l(m+\frac{1}{2})$ or
$\hbar\omega_r(n+\frac{1}{2}+\epsilon)$ $(n,m=0,1,2,\cdots)$, the
eigenvalues of the quadratic potentials of the wells. Furthermore,
the fact $D_k(\sqrt{2}y)=e^{-y^2/2}H_k(y)/(\sqrt{2})^k$ for a
non-negative integer $k$, implies that, if the eigenvalue is close
to $\hbar\omega_r(n+\frac{1}{2}+\epsilon)$
($\hbar\omega_l(m+\frac{1}{2})$), the eigenfunction of the
double-well system must be closely described by
$\psi_n^{ho}(a;l_r;x)$ ($\psi_m^{ho}(-b;l_l;x)$) around the bottom
of the right (left) well, with  an eigenfunction of a simple
harmonic oscillator $\psi_k^{ho}(c;l;x)$ ($\equiv H_k(\frac{x-c}{l})\exp[-\frac{(x-c)^2}{2l^2}]/\sqrt{\sqrt{\pi}l 2^k
k!}$).

\section{Two different regimes}

The analysis of the previous section shows that an eigenfunction of
the low lying excited states in the large separation limit has
significant amplitude either in both wells or in just one of the wells.
In this section, we will show that the eigenfunction of significant
amplitude in both wells must be accompanied by another eigenfunction
to form a doublet.  As in the symmetric case, a linear combination
of the doublet is responsible for the tunneling, and
we will calculate the tunneling visibility for the combination.
For the eigenfunctions localized in one of the wells, we will develop a
formula for the energy eigenvalue estimation.

\subsection{Tunneling visibility}

For the case that both $\mu$ and $\nu$ are close to integers $m$ and $n$,
respectively,
we define $\delta_\epsilon= \epsilon+n- \frac{\omega_l}{\omega_r}m$,
so that $|\delta_\epsilon|$ is equal or less than the minimum of
$\frac{1}{2}$ and $ \frac{\omega_l}{2\omega_r}$. In this case,
the corresponding eigenfunction gives
considerable probabilities in both of the left and right wells,
and $\mu$ and $\nu$ should be written as $\mu=m+\delta_\mu$ and
$\nu=n+\delta_\nu$ with $|\delta_\mu|,$ $|\delta_\nu|,$
$|\delta_\epsilon|\ll 1.$ From the fact that
$\delta_\mu=\frac{\omega_r}{\omega_l}(\delta_\nu+\delta_\epsilon)$,
Eqs.~(\ref{integer nu}) and (\ref{integer mu}) then yield
$\delta_\nu^2+\delta_\epsilon\delta_\nu-\delta_a^2=0,$ with
\begin{eqnarray}
\delta_a&=&\sqrt{\frac{\omega_l}{\omega_r}}\sqrt{\frac{1}{2\pi n!m!}}
  \left(\frac{n+\frac{1}{2}}{e}\right)^{\frac{2n+1}{4}}\cr
&&~~~~ \times  \left( \frac{m+\frac{1}{2}}{e}\right)^{\frac{2m+1}{4}}
 \exp\left[- \int_{-b_m}^{a_n} \frac{p(y)}{\hbar}dy\right].~~~
\end{eqnarray}

With
\begin{equation}
\delta_\pm= \frac{1}{2}(-\delta_{\epsilon}  \pm
\sqrt{\delta_\epsilon^2+4\delta_a^2}),
\end{equation}
when $\delta_\nu=\delta_-$,
the eigenfunction is written as
\begin{equation}
\psi^-(x)=
\frac{(-1)^n\delta_a
\psi_{m}^{ho}
(-b;l_l;x)+\delta_+\psi_{n}^{ho}(a;l_r;x)}{\sqrt{\delta_a^2+\delta_+^2}},
\end{equation}
while the eigenfunction of $\delta_\nu=\delta_+$ is
\begin{equation}
\psi^+(x)=
\frac{-(-1)^{n}\delta_+ \psi_{m}^{ho}
(-b;l_l;x)+\delta_a\psi_{n}^{ho}(a;l_r;x)}{\sqrt{\delta_a^2+\delta_+^2}}.
\end{equation}
The formal expression of  $\delta_{\pm}$ in terms of
$\delta_{\epsilon}$ and $\delta_a$  can be understood from the fact
that, for $|\delta_\epsilon|\ll 1$, the tunneling dynamics is
essentially described by that of a two-level system(TLS)
\cite{CDL,RMP,GDJH}. If  $\psi_n^{ho}(a;l_r;x)$ is written as a linear
combination of $\psi^-(x)$ and $\psi^+(x)$, the visibility of the
linear combination is
\begin{equation}
{\cal V}=1/[1+\frac{1}{2}(\frac{\delta_\epsilon}{\delta_a})^2].
\label{visibility}
\end{equation}

For the visibility estimation in Fig.~\ref{quartic}(c) and (d), the parameter is
determined from the consideration that, when $\delta_\epsilon=0$, the
energy splitting is given as $2\hbar\omega_r\delta_a$.
The fact that the tunneling dynamics significantly takes place for
$|\delta_\epsilon|\ll 1$, even with $m\neq n$, may be closely
related to the resonant enhancement of tunneling in the
multiple-well structures \cite{WM,Sias}. As in the BEC loaded into
an asymmetric double-well potential \cite{HWAHS}, if there are
noninteracting $N$ atoms in the ground state of $V(x)$ of small
$\delta_\epsilon$ ($\ll \delta_a$), the number difference between
the left and right wells is proportional to $\delta_\epsilon$.

\subsection{Energy eigenvalue estimation}

For an eigenfunction $\psi(x)$ of the system of $V(x)$ with the
energy eigenvalue $E$,   we have the identity
\begin{eqnarray}
&&E-\hbar\omega(n+\frac{1}{2})\cr
&&=\frac{\int_{-\infty}^{\infty}\left[ V(x)- V_r^{ho}(x)\right]
\psi(x)\psi_{n}^{ho}(a;l_r;x)
dx}{\int_{-\infty}^{\infty}\psi(x)\psi_{n}^{ho}(a;l_r;x) dx},~~~~~~~
\label{harmonic identity}
\end{eqnarray}
where $V_r^{ho}(x)= \frac{\hbar\omega_r}{2}\frac{(x-a)^2}{l_r^2}$.
In numerical calculations, this identity may be efficiently used
in estimating the energy eigenvalue of an eigenfunction which is close to
 $\psi_{n}^{ho}(a;l_r;x)$.

As the visibility also implies, when $\delta_\epsilon$ is much
larger than $\delta_0$, $\psi(x)$ of $E\approx \hbar
w_r(n+\epsilon+\frac{1}{2})$ is mostly localized in the right well,
and  around the bottom it will be closely described by
$\psi_{n}^{ho}(a;l_r;x)$, to give $\psi_{n}^{app}
(l_r;x)$ an approximation of $\psi(x)$ in this well.
In the other regions, Eqs.~(5,7) and the
WKB method can be used to find $\psi_{n}^{app}
(l_r;x)$. Eq.~(\ref{harmonic identity}) then may be
used to find a correction to $\nu$, as
\begin{eqnarray}
&&\nu-n +\epsilon \cr &&\approx\frac{\int_{-\infty}^{\infty}[ V(x)-
V_r^{ho}(x)]  \psi_n^{app}(l_r;x)\psi_{n}^{ho}(a;l_r;x) dx}
{\hbar\omega_r\int_{-\infty}^{\infty}
\psi_n^{app}(x)\psi_{n}^{ho}(a;x) dx}.~~~~~~
\end{eqnarray}
This approximation of a localized eigenfunction (ALE) can also be made similarly, for the eigenstate
of $E\approx \hbar w_l(m+\frac{1}{2})$ which describes  a probability distribution mostly localized
in the left well.

\section{ precision test: Asymmetric Double Oscillator}

In application of the WKB method for a symmetric double-well potential, it is
known that the energy splitting could be found accurately, if the (ground state)
energy eigenvalue
and thus the turning points are appropriately chosen \cite{WKB2}. If a well is quadratic with
angular frequency $\omega$, then $(j+\frac{1}{2})\hbar\omega$ ($j$: nonnegative integer) may be
a good estimation for an energy eigenvalue.

\begin{figure}
\includegraphics{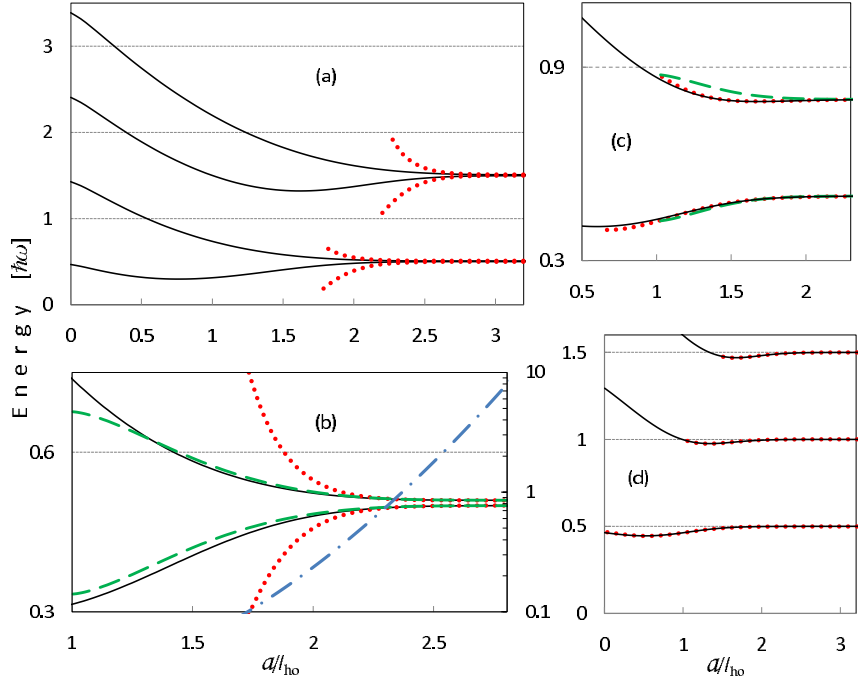}
\caption{(Color online) Energy eigenvalues for some lowest
eigenstates of the system of $V_D(\epsilon;x)$.  $\epsilon$ is
chosen as 0.01  for  (a) and (b), 0.3 for (c), and 0.5 for (d).
Calculated values (Solid lines); estimated values through the ALEs
(dotted lines) and through the approximation to a TLS (dashed
lines). For the ALEs of  the ground and second excited states,
Eq.~(\ref{integer mu}) and the turning points $a_n$, $b_n$
satisfying $V_D(\epsilon;a_n)=V_D(\epsilon;-b_n)=
(n+\frac{1}{2})\hbar\omega$  are used with $n=0$ and $n=1$,
respectively; and for the first and third excited states,
Eq.~(\ref{integer nu}) and $a_n$, $b_n$ of
$V_D(\epsilon;a_n)=V_D(\epsilon;-b_n)=
(n+\epsilon+\frac{1}{2})\hbar\omega$  are used with $n=0$ and $n=1$,
respectively. In the approximation to a TLS, $a_0$ and $b_0$ are
determined from  $V_D(0;a_0)=V_D(0;-b_0)= \frac{1}{2}\hbar\omega$.}
\label{Double Osc}
\end{figure}

In order to check the accuracy of the formalism
we have provided, avoiding the turning-point problem as much as possible,
we consider the system of the asymmetric double oscillator potential \cite{Song}
\begin{equation}
V_D(\epsilon;x)=\left\{\begin{array}{ll}
          \hbar\omega\left(\frac{x+\sqrt{a^2+2\epsilon l_{h0}^2}}
                   { \sqrt{2} l_{ho}} \right)^2  &
              ~{\rm for}~ x< 0,\\
          \hbar\omega\left[\left(\frac{x-a}{ \sqrt{2} l_{ho}} \right)^2+\epsilon\right]
          & ~ {\rm for}~ x\geq 0.
         \end{array}\right.
\end{equation}
For this system, since both wells are exactly quadratic, the eigenfunctions
are described by the parabolic
cylinder functions  on both  sides of $x=0$ \cite{Song,Merzbacher} and the
continuities of the eigenfunction and its derivative at $x=0$ can be
used to find the eigenvalues $E_0^D(a),E_1^D(a),\cdots$. As in
Fig.~\ref{Double Osc}, the calculations indeed show that, when $a$ is
a few times of $l_{ho}$, the eigenvalues of the low lying excited
states are close to $\hbar\omega(n+\epsilon+\frac{1}{2})$ or
$\hbar\omega(m+\frac{1}{2})$. To expose that the estimation through
the ALE is not valid when $\hbar\omega\delta_\epsilon$  is order of
or smaller than the energy difference of the adjacent energy
eigenstates, we add the ratio
$2\epsilon\hbar\omega/(E_1^D(a)-E_0^D(a))$ (dotted and dashed line)
in Fig.~\ref{Double Osc}(b).

When $\epsilon$ is as large as 0.3, the ALE gives
better results than the approximation to a TLS practically in the
whole range where both methods are applicable [Fig.~\ref{Double Osc}(c)].
As the approximation to a TLS is suggested by the WKB method,
Fig.~\ref{Double Osc} indeed shows that {\em WKB description} could be
{\em very accurate}. Fig.~\ref{quartic}(c) and (d) also suggests that
this accuracy is not limited to the systems of the wells
which are exactly quadratic.
This with the reasons the WKB method provides implies that,
if the potential $V(x)$ is quadratic until it reaches several times
of the zero-point energies $\hbar\omega_l/2$ and $\hbar\omega_r/2$
from the bottoms of the left
and right well, respectively, the energy eigenvalues of the low
lying excited states of the system must be close to the eigenvalues
of the quadratic potentials.

\section{conclusions and outlook}

We have shown, through the WKB method of quadratic connection formula,
that the systems of asymmetric double-well potentials
can be classified into two different regimes. In the regime of
eigenfunctions giving significant amplitude in both wells, the tunneling
dynamics could take place, while there is no tunneling
in the regime of localized eigenfunctions.  In this respect, the systems of
the eigenfunctions mostly localized in just one of the wells are very different
from those of the symmetric potentials. As  Fig.~\ref{quartic}(c), (d) and
Fig.~\ref{Double Osc} clearly show, the WKB description could be very accurate,
and the results given here may be valid for a system of the potential
whose wells are not exactly quadratic.

For the regime of localized eigenfunctions, even in the deep quantum limit,
it may be possible to confine a large number of noninteracting bosons
in just one of the wells.
For single-component fermions, in the light of the particle density
$\rho_R(\epsilon;x,t)=\sum_i\psi_{Ri}^*(\epsilon;x,t)\psi_{Ri}(\epsilon;x,t)$
(see, e.g., Ref.~\cite{BB}),
the number of fermions which can be confined in one of the wells
is limited by that of the localized eigenfunctions $\psi_{Ri}(\epsilon;x,t)$.
For the system of particles confined in just one of the wells, the
tunneling dynamics can be initiated and controlled by adjusting $\epsilon$
the potential energy difference between the bottoms of the double well, since,
if we change $\epsilon$ so that $\delta_\epsilon \ll 1,$
$\psi_{Ri}(\epsilon;x,t)$ turns into a linear combination of the
eigenfunctions of the new system.

In the periodic arrangement of double-wells of the optical lattice
where the tunneling is accompanied by a precession of the atom's
angular momentum \cite{HADGJ,KSSTO}, a  considerable time-periodic
{\em fluctuation of the population} of atoms in a spin state could
imply that the atomic spinors are in the states of {\em high
visibility}. Since $\delta_a$ is very small in the large separation limit
and the period of tunneling is inversely proportional to $\delta_a$,
if a tunneling phenomenon can be established over a long period of time,
it can be used for precision measurements.

\acknowledgments
The author thanks Professors Kyungwon An and Yong-il Shin for discussions
on experimental aspects.

\end{document}